\documentclass[11pt,titlepage,a4paper]{article}
	\usepackage{amsfonts}

\newtheorem{Prop}{Proposition}[section]		
\renewcommand{\theequation}{\thesection.\arabic{equation}}	
\newcommand{\QED}[1][6]{\hbox{\vrule width #1pt height #1pt depth 0pt}} 
\newcommand{\openone}{\leavevmode\hbox{\small1\kern-3.8pt\normalsize1}}
\newlength{\bo}\newlength{\ho}\newlength{\up}\newlength{\down}       
\newlength{\middle}                                                  
\newcommand{\bozho}{\leavevmode\hbox{\slshape\bfseries               
\settowidth{\bo}{BO}\settowidth{\ho}{HO}\settowidth{\middle}{/}
\settoheight{\up}{BOZHO}\settodepth{\down}{/}
\addtolength{\up}{+0.15\up}\addtolength{\bo}{+\middle}
\rule[\up]{\bo}{0.15ex}\hspace{-\bo}BO
\hspace{+0.09em}\raisebox{+0.17\up}{/}
\hspace{-0.20em}\raisebox{+0.71\up}{$\bullet$}
\hspace{-0.33em}\hspace{-1.14\middle}\raisebox{-0.4\up}{$\bullet$}
\hspace{-0.30em}\addtolength{\down}{-0.41\down}
\addtolength{\ho}{+1.5\middle}\rule[-\down]{\ho}{0.15ex}
\addtolength{\ho}{-\middle}\hspace{-\ho}\hspace{+0.18em}
\raisebox{+0.17\up}{HO}}}                                            
\newcommand{\BOZHO}
{\bozho$^{^{\text{\textregistered}\,} \text{\texttrademark} }$}      

\title{\vspace*{-1.23456789in}
\vspace*{-7ex}
{
\begin{flushright}
	  \textbf{\large LANL xxx E-print archive No. gr-qc/9806062}\\[5ex]
\end{flushright}
}
\huge IS THE PRINCIPLE OF EQUIVALENCE\\\vspace{6pt}  A PRINCIPLE?}
\vspace{10pt}
\author{
Bozhidar Z. Iliev
\thanks{Department Mathematical Modeling,
Institute for Nuclear Research and \mbox{Nuclear} Energy,
Bulgarian Academy of Sciences,
Boul. Tzarigradsko chauss\'ee~72, 1784 Sofia, Bulgaria}
$ $\thanks{E-mail address: bozho@inrne.acad.bg}
\thanks{URL: http://www.inrne.acad.bg/mathmod/bozhome/bozhome.htm}
}

\date{
 Ended: December 30, 1995\\
 Revised: April 19, May 2, October 24, 1996; February 12, 1997\\[4pt]
 Updated: June 13, 1998\\[4pt]
 Produced on: \fbox{\today} \\[4pt]
 LANL xxx E-print archive No. gr-qc/9806062\\[6pt]
 Sent for submission to J. Geom. Phys.: November 21, 1996\\
 Sent for resubmission to J. Geom. Phys.: March 3, 1997\\[3pt]
 Published in Journal of Geometry and Physics:\\
	vol.~24, No.~3, pp.~209--222, 1998\\[7ex]
 {\Huge\bozho}\\
\vspace{48pt}
\emph{
Locally inertial frames; Derivations; Linear connections
\\\vspace{8pt} \textbf{1996 PACS numbers}:\\
02.40.Hw, 03.30.+p, 04.20.Cv, 04.50.+h%
, 02.40.Ky, 02.40.Sf
\\\vspace{8pt} \textbf{1991 MSC numbers}:\\
53B99, 53B50,
83D05, 83C99, 83E99,
53B05, 53B15 
}
}

\begin{document}

\renewcommand{\thefootnote}{\fnsymbol{footnote}}
\maketitle			
\renewcommand{\thefootnote}{\arabic{footnote}}

\tableofcontents

\begin{abstract}

	The work argues the  principle of equivalence to be a theorem and not
a principle (in a sense of an axiom). It contains a detailed analysis of the
concepts of normal and  inertial frame of reference. The equivalence principle
is proved to be valid (at every point and along every path) in any
gravitational \mbox{theory} based on linear connections. Possible
generalizations of the equivalence principle are pointed out.

\end{abstract}

\section {\bfseries Introduction}
\label{I}
\setcounter {equation} {0}

The principle of equivalence played an important role at the
early stages of development of general
relativity~\cite{MTW,Weinberg,Fock,JL-Anderson,R-Torretti}.
Now, despite historical positions, it is often mentioned as a procedure
for transferring results from flat space-time(s) to curved
one(s)~\cite[ch.~16]{MTW}.  Mathematically this is reflected in the minimal
coupling principle used to transfer the Lagrangian forma\-lism from flat to
curved manifolds by replacing the flat metric with the
(pseu\-do\mbox{-)}Riemannian one and the usual (partial) derivatives with
covariant ones~\cite{Heyde}.

	The equivalence principle is almost everywhere considered as a
statement that can't be proved or need not to be proved as it is
`evident' from certain positions and whose consequences are `reasonable
enough' to be taken as a true~\cite{JL-Anderson,R-Torretti}.

	The present paper asserts the opinion that when the mathematical
background of a gravitational theory is chosen, then the (strong) equivalence
`principle' becomes a theorem (true or not) that can be proved. This is in
accordance with the conclusions of~\cite[\S~61]{Fock}. There is another case
when the equivalence principle is used for selecting the mathematical
structure of a gravitational theory. In this case it acts primarily as a
principle (axiom), but after this selection is made, it again, becomes a
theorem.

	In~\cite{Synge} (see also~\cite[pp.~5, 160]{Mitskevich})
is recognized the historical role of the equivalence principle in general
relativity but its exact contents and importance are put under question. By
our opinion the latter is a consequence of (some of) the indistinct
formulations of this principle and the problem is ---
is it a theorem or an axiom?
These problems are solved completely in the present work.  That takes off
some of Synge's questions. But we do not share his mind that the
equivalence principle is not important nowadays. He is right that now
general relativity can be formulated without it. But general relativity
is compatible (consistent) with the equivalence principle
(in a sense that in this theory it is a provable theorem)
as it must be because this principle reflects important empirical observation.
Besides, the significance of the principle of equivalence arises (maybe
implicitly) in any new gravitational theory as only theories compatible with
it can survive.

	The present investigation concentrates mainly on the mathematical
aspects of the equivalence principle. A physical discussion of this principle
can be found
in~\cite[see in particular pp.~133--137]{R-Torretti},
~\cite[pp.~334--338]{JL-Anderson}, or in~\cite[\S\S~8.2, 9.6]{Moller}.

	This work is based mainly
on~\cite{f-Frames-n+point,f-Bases-n+point,f-Bases-path,f-Bases-general}
and is organized as follows. Sect.~\ref{II} is a brief review of the
equivalence principle and its mathematical formulation. Sect.~\ref{III} is
devoted to some mathematical theorems closely connected to the subject of
this article.  Physical conclusions from them are made in Sect.~\ref{IV}.
Sect.~\ref{V} contains remarks about possible extensions of the area of
validity of the equivalence principle. \ref{A} reviews and
discusses some terminological problems. \ref{B} contains certain
results concerning derivations. \ref{C} outlines the proofs of
propositions used in this work.

\section {\bfseries The equivalence principle from \\
physical and mathematical point of view}
\label{II} \setcounter {equation} {0}

	 Different formulations of the equivalence
principle can be found. They state in one or the other form that (at a
point) in a suitable frame of refe\-rence the laws of special and general
relativity coincide:  In~\cite[ch.~16]{MTW} it reads: ``In any local
Lorentz frame at any time and place in the Universe all (nongravitational)
physical laws take their special relativity form''. In~\cite{Weinberg}, one
finds it as the assertion that at any space-time point in arbitrary
gra\-vitational field a ``locally-inertial coordinate system can be chosen,
in which in a sufficiently small neighborhood of the point, the Nature laws
will have the same form as in non-accelerated Cartesian coordinate systems''.
In~\cite{Heyde} it states that ``locally the properties of special
relativistic matter in a non\-inertial frame of reference cannot be
distinguished from the properties of the same matter in a corresponding
gravitational field''. In~\cite[\S~9.6]{Moller} the equivalence principle is
formulated as follows: ``at any point all Nature laws, expressed in local
Lorentz coordinates, have the same form as in special relativity''.

	In fact, these are formulations of the \emph{strong} equivalence
principle which is discussed, for instance, in
~\cite[Sect. 5.2]{R-Torretti} (see also the references therein).
The several weak forms of the principle of equivalence are not a subject of
the present investigation.

	Above, as well as in other `physical' publications, the concepts
`local' and `locally' are not well defined from mathematical view-point
and often mean an ``infinitesimal surrounding of a fixed point of
space-time''~\cite{Heyde}. Their strict meaning may be at a point, along a
path (curve), in a neighborhood, or on another submanifold (or, generally,
subset) of space-time. Below we will have in mind just this, every
time specifying the particular situation.

	As we saw above, in the equivalence principle is involved a special
class of coordinate systems or frames (of reference), usually called (local)
inertial~\cite{Weinberg} or (local) Lorentz~\cite{MTW}
in the physical literature and normal
(and, by some authors, geodesic or Riemannian) in the mathematical
one~\cite[ch.~V, Sect.~3]{Schouten/physics} (see \ref{A}). The main
property of a frame of this class is that in it one can `locally' neglect
the effects of gravity (or of the accelerated motion of the frame), or, more
strictly, that in it the gravitational field strength is `locally'
transformed to zero (or vanishes). Mathematically this is the corner-stone
of the equivalence principle: if such frames do not exist it is meaningless
and conversely, if they exist it is meaningful, and the problem whether the
equivalence principle is a principle (an axiom) or a theorem depends on the
approach to the concrete theory under consideration (see Sect.~\ref{IV}).

	In all of the gravitational theories known to the author the
gravitational field strength  is locally identified with the components of a
certain linear connection, for instance with the Christoffel symbols formed
from the metric  (Levi-Civita's connection) in general
relativity~\cite{MTW,Weinberg} or with the coefficients of the Riemann-Cartan
connection in the  $U_4$ theory~\cite{Heyde}. Just this point connects
physics with mathematics here and makes it possible the strict mathematical
consideration of the above problem. In fact, in this context, the above
special frames are coordinate systems (or local bases) in which the
components  of the corresponding linear connection locally vanish.

	So, if locally the gravitational field strength is identified with
the local components of a linear connection  $\nabla$, then it is meaningful
to be spoken about the equivalence principle on some subset  $U$ of the
space-time  $M$ if and only if in (a neighborhood of)  $U$ exist frames
(coordinates, bases) in which the connection's components vanish on  $U$.
Thus there arises the mathematical problem for finding, if any, the linear
connections  $\nabla$ on the set  $U$ and the coordinates, called
\textit{normal}, in a neighborhood of $U$ in which  the components of
$\nabla$ vanish on  $U$. To the author are known the following basic results
on this problem.

	According to~\cite[ch.~V, Sect.~3]{Schouten/physics} the existence of
normal coordinates at a point ($U=\{x_0\},\ x_0\in M$) for symmetric linear
connection has been proved at first in~\cite{Veblen}. In 1922
Fermi~\cite{Fermi} has proved the existence of normal coordinates along any
curve without self-intersections in the pseudo-Riemannian mani\-fold of the
general relativity. In many textbooks
(see, e.g.,~\cite{Schouten/Ricci,Schouten/physics}) it is
proved that for symmetric linear connections normal coordinates
exist in a neighborhood iff the connection is flat in it. The general case
for symmetric linear connections is investigated in~\cite{ORai} where
necessary and sufficient conditions for the existence of normal coordinates
on submanifolds were found. All these results concern torsion free,
i.e. symmetric, linear connections. In the corresponding works it is also
mentioned that for nonsymmetric linear connections there are no
normal coordinates (more precisely, holonomic normal coordinates don't
exist). It seems that in~\cite{Heyde},
in fact without proof, the existence of anholonomic normal coordinates at a
point for nonsymmetric linear connections was mentioned
for the first time. In 1992,
in~\cite{f-Bases-n+point} and in~\cite{f-Bases-path} the existence
of generally anholonomic local normal coordinates, called
there special bases, was proved at a point and along a path,
respectively, not only for any linear connection but also for arbitrary
derivations of the tensor algebra over a differentiable manifold. The
work~\cite{f-Bases-n+point} among others deals with the problem in a
neighborhood: the sought for (anholonomic) normal coordinates exist only
in the flat case (zero curvature of the derivation or connection). The
paper~\cite{f-Bases-general} contains necessary and/or sufficient
conditions for existence, holonomicity, and uniqueness of normal
coordinates (special
bases) on sufficiently general subsets of a differentiable manifold for
arbitrary derivations of the tensor algebra over it that, in particular,
may be linear connections. In 1995 in~\cite{Hartley} (independently
of~\cite{f-Bases-n+point}) the existence of anholonomic normal
coordinates (frames) at a point was proved for linear connections with
torsion, a result which is a very special case of the ones
of~\cite{f-Bases-general} or~\cite{f-Bases-n+point}.

	The cited results, some of which will be discussed in the next
section, are the strict mathematical base for analyzing the equivalence
principle.

\section {\bfseries On the general existence of normal coordinates}
\label{III}
\setcounter {equation} {0}

	As we have said in the previous section, the problems connected with
the existence (and uniqueness) of normal coordinates for symmetric linear
connections were more or less completely investigated
in~\cite{Veblen,Fermi,ORai}.
In~\cite{f-Frames-n+point,f-Bases-n+point,f-Bases-path,f-Bases-general}
analogous problems were studied in the case of arbitrary derivations of
the tensor algebra over a differentiable manifold. In particular these
derivations can be covariant differentiations (linear connections) with or
without torsion. Thus, these works, a brief review of which is presented
below, incorporate as their special cases the above cited ones concerning
torsion free linear connections.

	Any (S-)derivation of the tensor algebra over a manifold  $M$ is a
map $D:X\mapsto D_X=L_X+S_X$, where  $X$ is a vector field,  $L_X$ is the Lie
derivative along  $X$, and  $S_X$ is (depending on  $X$) tensor field of type
$(1,1)$ considered here as derivation~\cite{K&N-1,f-Bases-n+point}.

	If  $\{E_i\}$ is  a field of vector bases in the tangent
bundle to $M$, then the \textit{coefficients} $(W_X)_{j}^{i}$ of
$D$ are defined, e.g., through
\begin{equation}
D_X(E_j) = (W_X)_{j}^{i}E_i.               	\label{3.1}
\end{equation}
\noindent
Here and below all Latin indices run from 1 to $\dim(M)$ and summation from 1
to  $\dim(M)$ over repeated indices on different levels is assumed.

	Let  $W_X:=[ (W_X)_{j}^{i} ]$ be the matrix formed from the
coefficients $(W_X)_{j}^{i}$ of  $D$. The change
$\{E_i\}\to\{E^\prime_i:=A_{i}^{j}E_j\}$ of the basic vector fields induces
\begin{equation}				\label{3.2}
W_X\to W_{X}^{\prime}=A^{-1}(W_X A + X(A))
\end{equation}
\noindent
with  $ A := [A_{j}^{i}]$ and $X(A)$ being the action of $X$ on  $A$, i.e.
$ X(A) = [ X(A_{j}^{i} ] = [ X^kE_k(A_{j}^{i}) ] $.

	From~(\ref{3.1}) or~(\ref{3.2}) it is evident that  $D$ is a
covariant differentiation $\nabla$ with (local) coefficients
$\Gamma_{jk}^{i}$  in
$\{E_i\}$ iff  $(W_X)_{j}^{i}=\Gamma_{jk}^{i}X^k$, i.e. if  $W_X$ depends
linearly on  $X$. In general, $D$ is said to be \textit{linear on (in)
$U\subseteq M$ or along a map $\eta:Q\to M$} for some set $Q$ if in some basis
(and hence in \textit{all} bases)  $\{E_i\}$ the relation
$W_X(x)=\Gamma_k(x)X^k(x)$ is fulfilled for some matrix functions $\Gamma_k$
and $x\in U$ or $x\in\eta(Q)$ respectively.

	The (operators of) \textit{curvature} $R^D$ and \textit{torsion} $T^D$
of a derivation  $D$ are, respectively,
$R^D(X,Y):= D_X\circ D_Y - D_Y\circ D_X -D_{[X,Y]}$
and
$T^D(X,Y):= D_XY -D_YX -[X,Y]$
for any vector fields $X$ and $Y$, $[X,Y]$ being their commutator.

	Now the problem interesting for us has the following formulation.
Let there be given a subset $U\subseteq M$. There have to be found all
derivations $D$ and the corresponding fields of bases $\{E_i\}$, defined on
$U$ or on a neighborhood of $U$, in which the components of  $D$ vanish on
$U$, i.e.  $W_X(x)=0$ for $x\in U$.  If such bases
(frames) exist, we call them \textit{normal bases} (resp.
\textit{normal frames}) for $D$ (on  $U$).  Here and below we prefer to speak
about normal bases (or frames) instead of normal coordinates because these
bases (frames) are generally anholonomic, i.e. in the usual sense (holonomic
or integrable) coordinates with the needed property do not exist and one has
every time, when mentioning them, to add the appropriate adjective
`anholonomic' or `holonomic'.

	Now we shall present some basic results
from~\cite{f-Frames-n+point, f-Bases-n+point, f-Bases-path, f-Bases-general}
concerning the existence, uniqueness, and holonomicity of normal frames.

	In neighborhoods the following results are
valid~\cite{f-Frames-n+point,f-Bases-n+point}:

\begin{Prop}	\label{Prop3.1}
	In a neighborhood $U\subseteq M$ there exist a normal frame for a
derivation $D$ if and only if it is a flat linear connection or iff it
is flat ($R^D=0$) and $\left.D_X\right|_{X=0}=0$ in $U$.
\end{Prop}

\begin{Prop}	\label{Prop3.2}
	The normal bases in $U$ for $D$, if any, are connected with
(homogeneous) linear transformations with constant coefficients and are
ho\-lonomic (anholonomic) iff $T^D =0$ (resp. $T^D\not =0$) in $U$.
	\end{Prop}

	Hence the flat (in  $U$) linear connections are the only derivations
for which there exist normal bases in neighborhoods. These frames are
holonomic iff the connection is symmetric (torsion free).

	At a given point our problem is solved
by~\cite{f-Frames-n+point, f-Bases-n+point}:

\begin{Prop}	\label{Prop3.3}
	At a point $x_0\in M$ there exists a normal frame for a derivation
$D$ iff  $D$ is linear at $x_0$.
\end{Prop}

\begin{Prop}	\label{Prop3.4}
The normal bases for $D$ at $x_0$, if any, are connected by linear
transformations whose matrices vanish at $x_0$ under the action of the normal
basic fields, and they are holonomic iff  $D$ is torsion free at  $x_0$.
\end{Prop}

	As a linear connection is, evidently, a linear at (every)  $x_0$
derivation, the last two propositions contain as a special case the
hypothesis formulated in~\cite{Heyde}, as well as its strict formulation and
proof in~\cite{Hartley}: any linear connection admits normal frames at every
fixed point which are holonomic iff it is symmetric.

	Along an arbitrary path  $\gamma:J\to M$,  $J$ being a real interval,
the following propositions are fulfilled~\cite{f-Bases-path}:

\begin{Prop}	\label{Prop3.5}
	Along $\gamma$ (i.e. on $\gamma(J)$) there exists a normal basis for a
derivation  $D$ iff $D$ is linear along  $\gamma$ (i.e. on $\gamma(J)$).
\end{Prop}

\begin{Prop}	\label{Prop3.6}
	The normal along  $\gamma$ bases for $D$, if any, are connected
through linear transformations whose matrices vanish along  $\gamma$ under
the action of the normal basic fields. If they are holonomic, then
$D$ is torsion free on  $\gamma(J)$ and, conversely, if $D$ is torsion free
on  $\gamma(J)$ and there is a smooth normal basis along  $\gamma$, then
all of them are holonomic.
\end{Prop}

	As a linear connection $\nabla$ is a derivation linear along any path,
we see that any linear connection admits normal frames along every fixed path.
If there is a holonomic basis for $\nabla$ normal along  $\gamma$, then
$\nabla$ is symmetric and if  $\nabla$ is symmetric, and there is a normal
basis for it which is smooth along  $\gamma$, then all such bases are
holonomic.  In particular, for symmetric  $\nabla$ and paths without
self-intersections we get in this way the classical result of~\cite{Fermi}.

	If one is interested of derivations along paths (see the definition
in~\cite[Sect. III]{f-Bases-path}), there always exist holonomic as well as
anholonomic normal bases along any path $\gamma$. In particular, this is
true for the covariant differentiation $\nabla_{\dot\gamma}$ along $\gamma$
corresponding to a linear connection $\nabla$ ($\dot\gamma$ is the
tangent to  $\gamma$ vector field).

	The general situation concerning normal bases is the
following~\cite{f-Bases-general}:

	\begin{Prop}	\label{Prop3.7}
	If on the set $U\subseteq M$ there exists a normal basis for a
derivation  $D$, then  $D$ is linear on $U$.
	\end{Prop}

	But the opposite to this proposition is generally not valid (cf.,
e.g., proposition~\ref{Prop3.1}).

	\begin{Prop}	\label{Prop3.8}
	In a set  $U$ the normal bases for $D$, if any, are connected by
linear transformations whose matrices vanish on  $U$ under the action of
these normal basic vector fields. If there is such a holonomic basis, then
$D$ is torsion free on $U$ and, conversely, if  $D$ is torsion free on $U$
and there is in $U$ a smooth normal basis for $D$, then all normal in $U$
bases for $D$ are holonomic.
	\end{Prop}

	The theorem 4 of~\cite{f-Bases-general} expresses a necessary
and sufficient condition for exis\-tence of normal bases (frames) for
linear derivations along maps with separable points of
self-intersection. In particular it covers the case of arbitrary
submanifolds of the space-time and the case of arbitrary linear
connections, thus generalizing the results of~\cite{ORai}. Here we shall
mention only the following corollary of this theorem. The zero- and
one-dimensional cases are the only ones in which normal frame always exist
for derivations which are linear on the corresponding sets (see
resp. propositions~\ref{Prop3.3} and~\ref{Prop3.5}). In particular this
is true for linear connections. On submanifolds of dimension
$p=2,\ldots,\dim M$ (for $\dim M\geq2$) normal frames exist only as an
exception in a case when some conditions are fulfilled (for $p=\dim M$
cf. proposition~\ref{3.1}).

\section {\bfseries The equivalence principle: Axiom or a theorem?}
\label{IV}
\setcounter{equation} {0}

	It was shown in Sect.~\ref{II} that the equivalence principle is
meaningless without a clear and strict understanding of what is a local
inertial frame. Physically it can be defined as a frame in which
the gravitational field strength (locally) vanishes. But then the
question arises how this strength is described mathematically. In all
(non-quantum) gravitational theories known to the author the
gravitational field
strength is (locally) identified with the components of some linear
connection which leads to the identification of the class of inertial frames
with the class of normal frames for this linear connection. Hence, in these
theories the \emph{physical concept `inertial frame' coincides with the
mathematical concept `normal frame'}.
In this way also  automatically the problem of what `local' (or
`locally') strictly means in the equivalence principle is solved: it
simply means the set(s) on which the corresponding normal frame(s) is (are)
defined.

	The results of Sect.~\ref{III} imply that normal frames exist not
only for \mbox{linear} connections but also for more general derivations
(which are linear on the corresponding sets). So, the equivalence principle
can be formulated for theories in which the gravitational field strength is
identified with the components of certain derivation of the tensor algebra
over the space-time. In this case one has to identify the inertial and normal
frames too.

	If one wants the normal frames to exist not only on a particular set
(e.g. on a given path) but also on some class of subsets of the space-time
(e.g. on all paths), then he again arrives to the case of linear connections
if these subsets cover the whole space-time. (In the last case by
proposition~\ref{Prop3.7} the derivation is linear at any space-time point
which means that it is a linear connection.) Combining this results with
propositions~\ref{Prop3.3} and~\ref{Prop3.5} one derives:

\begin{Prop}	\label{Prop4.1}
The linear connections (covariant differentiations) are the only derivations
for which normal bases exist at every space-time point or/and along every
path in it.
\end{Prop}

	On other (families of) sets, even for linear connection, normal
frames exist only as an exception (see, e.g., proposition~\ref{Prop3.1}
and~\cite{f-Bases-general}).

	Consequently, if one tries to formulate the equivalence principle he
has to suppose that the gravitational field strength is identified with the
coefficients of some linear connection. If this is done, then there exist
local inertial frames ($\equiv$  normal frames).

	Until now the `first part' of the equivalence principle was
discussed: it concerns inertial (normal) frames from mathematical point of
view. Its `second part' presupposes the existence of inertial frames and
states that in them the ``non-gravitational physical laws take their special
relativity form''. But here the question arises: when and in which frames
the special relativity (and the physical laws in it) is (are) valid?

	The answer is: in frames which are not accelerated or in which the
gravitational field strength vanishes which, because of the empirical
equality between inertial and gravitational masses, is one and the same
thing~\cite{Einstein-Relativity}. Such frames are called, by definition,
inertial too. This is not accidental because their class coincides with the
above-considered class of normal frames in which the gravitational field
strength vanishes too. Hence, it turns out that by definition, empirically
based on the equality of inertial and gravitational masses,  the special
relativity Nature laws are valid in the inertial frames.

	So, what does the equivalence principle state in the end?
The existence
of iner\-tial frames? No, because they are needed for its formulation and the
fact of their existence  is a consequence of the theory's mathematical
background. Where the special relativity laws are valid? No, because
this is either a question of definition: once the special relativity laws are
established and experimentally checked, one has to extrapolate this fact by
mathematically describing where they are valid. The above discussion shows
that in this context
\textbf{\emph{%
the equivalence principle asserts the coincidence of the
two types of inertial frames: the normal frames, in which the components
of a linear connection (or some other derivation) vanish, and the
inertial frames, in which special relativity is valid.
}}%
But, as it was demonstrated
above, this is a consequence of the fact that the gravitational field
strength is mathematically described by the components of a certain linear
connection. Thus, from this positions,
\emph{\textbf{the equivalence principle is a theorem}}.

	It seems that for the first time such a conclusion was made
in~\cite[\S~61]{Fock} in the case of general relativity, where it is
asserted that the equivalence principle ``is contained in the
hypotheses of the Riemannian character of space-time and mathema\-tically is
expressed in the possible introduction of local geodesic (i.e. normal - B.I.)
coordinate systems along a time-like world line''~\cite[p.~307]{Fock}.

	Can the equivalence principle be considered as an axiom? Our opinion
is that this is also possible, but not in its usual formulation(s) (see
Sect.~\ref{II}). For this purpose the `equivalence principle' should be
formulated as follows:
\emph{%
in any local frame of reference the gravitational
field strength is described through identifying it with the local
coefficients in this frame of a certain li\-near connection
(or another derivation).
}%
Implicitly in this statement the equality between the inertial and
gravitational masses is incorporated which is supposed to be valid before
the formulation of the usual equivalence principle, which in its turn, as was
demonstrated above, is a consequence of it.

\section {\bfseries Can the equivalence principle be generalized?}
\label{V}
\setcounter {equation} {0}

	In the usual formulation(s) of the equivalence principle the
question for its generalization does not stand at all: it concerns a
single theory (general  relativity~\cite{MTW,Weinberg}) and its validity in
other theories (such as the $U_4$ gravity theory~\cite{Heyde}) was under
question until recently. Our investigation shows that it is meaningful also
in any gravitational theory based on linear connections. It is valid in such
a theory at every point and along any path. On other subsets of the
space-time it can be valid only as an exception. One can also formulate the
equivalence principle in gravitational theories based on derivations more
general than covariant differentiation. In such theories it can, in
general, be valid on particular subsets of  the space-time. If its validity
in them is demanded on the whole space-time, then with necessity the
corresponding derivation must be a covariant differentiation, i.e. one
arrives again at a theory based on linear connections.

	In sum,
\emph{\textbf{%
the equivalence principle (in its usual formulation(s)) is
valid in the whole space-time (at any point or along any path) in all
gravitational theories based on linear connections.
}}%
(Note that the new
formulation of the equivalence principle, presented at the end of the previous
section, serves just to select those theories.)

	Further generalizations of the equivalence principle are possible in
two directions: by generalizing the (mathematical) concept of `normal'
frame or by generalizing the description of the gravitational interaction (on
the base different from the one of linear connections).

	One possible such generalization is outlined
in~\cite{f-EqPr-LT-tensors}. In it one supposes the tangent to the
space-time bundle to be endowed with a linear transport along paths, which
may not to be a parallel transport assigned to a linear connection. (For the
general theory of such transports - see~\cite{f-LTP-general}.) The
gravitational field strength is then identified with the transport's
coefficients. (The gravitational field itself can be described through the
transport or its curvature.) Define the class of the normal frames to be the
one of all bases (frames) in which the transport's coefficients vanish along
an arbitrary given path. The so-defined normal frames always exist along any
path or at any point (which is a degenerate path). In such a gravitational
theory, which will be studied elsewhere, the equivalence principle is valid,
for instance, in any of its formulations given in Sect.~\ref{II}. Due to
the equivalence established in~\cite{f-LTP-general} between linear transports
along paths (generally in vector bundles) and derivations along paths, the
sketched base for a possible gravitational theory can be formulated
(equivalently) in terms of derivations along paths.
Evidently, in such terms it is a straightforward generalization of
the theories based on linear connections.

	Another way for generalizing the equivalence principle is to extend
the `physical' area of its validity, i.e. to apply it to fields different
from the gravitational one (cf.~\cite{Kapuszik}). The reason for such
possibility is the fact that the gauge (Yang-Mills) fields are from
mathematical view-point linear connections (on vector bundles). This suggests
the idea for such a formulation of the equivalence principle that it
concerns all fields (interactions) described by means of gauge theories.

\appendix
\renewcommand{\thesection}{Appendix~\Alph{section}}
\renewcommand{\theequation}{\Alph{section}.\arabic{equation}}

\section
[\hspace*{11.5ex}Normal, geodesic, Lorentz, and inertial frames]
{\bfseries Normal, geodesic, \\ Lorentz, and inertial frames}
\label{A}
\setcounter{equation} {0}

	We called normal a special kind of local bases, frames, or
coordinates investigated in the present paper. This needs some explanations.

	For symmetric linear connections the local coordinates in which their
components vanish at a given point are called normal
in~\cite[ch.~V, Sect.~3]{Schouten/physics} or in~\cite[\S~11.6]{MTW}.
In~\cite[ch.~III, \S~8]{K&N-1} and in~\cite[p.~278]{R-Torretti}
the local coordinates normal at a point, introduced there via the exponential
map, for any linear connection (symmetric or not) are defined as such for
which the symmetric part of the connection's components vanish at this point.
Evidently, the latter definition includes the former one as a special case.
Note that the both definitions originate from the consideration of the
equation of geodesic lines~\cite{Schouten/physics,K&N-1,R-Torretti}.
This is the primary reason to call these local coordinates geodesic (or
Riemannian, or normal Riemannian~\cite[\S~11.5]{MTW}) in the special case of
a Riemannian manifold~\cite[\S~42, p.~201]{Fock}, where they are (some times)
equivalently introduced via the condition that in them the partial
derivatives of the metric's components vanish at a given
point~\cite[\S~42]{Fock}.

	The case of a symmetric linear connection is investigated
in~\cite[ch.~III, \S~7, pp.~156--158]{Schouten/Ricci} (see the references
therein too). There is made a distinction between geodesic and normal at a
point local coordinates. Geodesic coordinates are called the ones in which at
that point vanish the connection's components and normal coordinates are
called the geodesic ones satisfying at the given point equation (7.23)
of~\cite[ch.~III, \S~7]{Schouten/Ricci} which, in particular, implies the
vanishing at that point of the connection's components together with their
symmetrized partial derivatives. (Note that the possibility for the existence
of the last type of coordinates is ensured by our (non-)uniqueness result
expressed by proposition~\ref{Prop3.4} with which is compatible the mentioned
equation.)  Analogous opinion is shared in~\cite[pp.~13--14]{Mitskevich}.

	It is known that the symmetric part of the connection symbols of
arbitrary linear connection $\nabla$  are directly connected with the
equation of geodesic lines (curves, paths) and uniquely determine
them~\cite{Schouten/Ricci,K&N-1}. By our opinion, this suggests the following
convenient convention. Call normal or resp. geodesic on a set~$U$  a local
coordinate system (basis, or frame), defined in a neighborhood of~$U$, in
which the local components of $\nabla$ or resp. their symmetric parts vanish
on~$U$. Thus in the torsion free case the concepts of normal and geodesic
coordinate system coincide. Generally a normal frame is geodesic, the
converse being not valid. In this sense, the normal coordinates described
in~\cite[p.~158]{Schouten/Ricci} are a special type of (our)
normal coordinates, specified by the additional conditions
described in this reference. These conditions are consistent with
proposition~\ref{Prop3.4}.  Note that the proposed definition is in
accordance with the special one used in~\cite{Hartley}.

	If one adopts the suggested convention, then the generalization from
linear connections to arbitrary derivations~$D$  of the tensor algebra over a
manifold is evident: only the concept of a normal frame is applicable because,
generally, of some symmetry properties of the coefficients of $D$ cannot be
spoken about. This explains the terminology accepted in the present paper.

	Let us mention that the so-defined normal bases for~$D$  in~$U$ have
a connection with a kind of generalized geodesic lines corresponding to~$D$
(cf.~\cite{f-LTP-appl}) which will be discussed elsewhere.

	In the physical literature, contrary to the mathematical one,
there is a unique understanding what local inertial and Lorentz frames
are. A local Lorentz coordinate system is defined for the
(pseudo-)Riemannian space-time of general relativity as a one in which at a
given point (or another set) the metric tensor coincides with the Minkowski
metric tensor and all partial derivatives of the metrical components are
zeros at this point (see, e.g.,~\cite[\S\S~8.5, 8.6, 13.6]{MTW}
or~\cite[\S~9.6]{Moller}). (Note that this definition admits an evident
generalization to arbitrary (pseudo-)Riemannian manifolds: only the
Minkowski metric tensor has to be replaced with arbitrary fixed tensor.) A
local inertial frame (of reference) at a given point (or another set) is
defined as a one in which at this point (or another set) the gravitational
effects (or more precisely, the gravitational field strength) vanish
(see~\cite[\S\S~1.3, 1.6]{MTW} or~\cite[\S~9.6]{Moller}). When the
gravitational field strength is identified with the local components of some
linear connection, which is the usual situation~\cite{MTW,Moller,Fock,Heyde},
this means the vanishing of the connection's components at the given point.
In general relativity this leads to the fact that any local Lorentz system is
a local inertial frame~\cite[\S~13.3]{MTW}.

	Thus, if the gravitational field strength is locally identified with
the local components of some derivation $D$, then only the concept of a local
inertial frame survives. Besides, if there is presented (may be independently)
a metric, then there arises also the class of local Lorentz frames; of such a
type are the metric-affine gravitational theories. Generally, these types of
frames, if both exist, need not be connected somehow with each other.

\section
[\hspace*{11.5ex} On derivations of the tensor algebra over\\ a manifold]
{\bfseries On derivations \\ of the tensor algebra over a manifold}
\label{B}
\setcounter{equation} {0}

	A derivation of the tensor algebra $\mathcal{T}(M)$ over a
differentiable manifold $M$  is a linear map
$D:\mathcal{T}(M)\to\mathcal{T}(M)$ which satisfies the Leibnitz
differentiation rule with respect to the tensor product, preserves the
tensor's type, and commutes with the contractions of the tensor
fields~\cite[ch.~I, \S~3]{K&N-1}. By~\cite[proposition 3.3 of chapter I]{K&N-1}
any $D$ admits a unique representation in the form $D=L_X+S$ for some (unique
for a given $D$) vector field $X$  and tensor field  $S$  of type $(1,1)$.
Here $S$  is considered a derivation of $\mathcal{T}(M)$~\cite{K&N-1}, which
for a covariant differentiation $\nabla$ is given through
 $S_X(Y)=\nabla_X(Y)-[X,Y]$, $Y$  being a vector field.

	Let $\{ E_i,\ i=1,\ldots,n:=\dim(M) \}$ be a (coordinate or
not~\cite{Schouten/Ricci}) local basis (frame) of vector fields in the
tangent to $M$  bundle. It is holonomic (anholonomic) if the vectors
$E_1,\ldots,E_n$ commute (do not commute)~\cite{Schouten/Ricci}. Let $T$ be a
$C^1$ tensor field of type $(p,q)$, $p$ and $q$  being integers or zero(s),
with local components
 $T_{j_1 \ldots j_q}^{i_1 \ldots i_p}$
with respect to the tensor basis associated with $\{E_i\}$. Here and below
all Latin indices, maybe with some super- or subscripts, run from $1$ to
$n:=\dim(M)$. Using the explicit action of  $L_X$  and $S_X$ on tensor
fields~\cite{K&N-1} and the usual summation rule about repeated indices
on different levels, we find the components of $D_XT$ to be
	\begin{eqnarray}
\nonumber
\left( D_X T \right) _{j_1 \ldots j_q}^{i_1 \ldots i_p}  & = &
X\left( T_{j_1 \ldots j_q}^{i_1 \ldots i_p} \right) +
\sum_{a=1}^{p}\left(W_X\right)_{k}^{i_a}
	T_{j_1 \ldots j_q}^{i_1 \ldots i_{a-1} k i_{a+1} \ldots i_p} \> - \\
\label{B.1}  & - & \>
\sum_{b=1}^{q}\left(W_X\right)_{j_b}^{k}
	T_{j_1 \ldots j_{b-1} k j_{b+1} \ldots j_q}^{i_1 \ldots i_p}.
	\end{eqnarray}
Here $X(f)$ denotes the action of  $X=X^iE_i$ on the $C^1$ scalar function
$f$, i.e.  $X(f)=X^kE_k(f)$, and  the explicit form of $W_X$
(cf.~(\ref{3.1})) is
 	\begin{equation}	\label{B.2}
 \left( W_X \right)_{j}^{i} =
 \left( S_X \right)_{j}^{i} - E_j(X^i) + C_{kj}^{i}X^k
 	\end{equation}
 where $C_{kj}^{i}$ define the commutators of the basic vector fields by
 $[E_j,E_k] =  C_{jk}^{i}E_i$.

	From~(\ref{B.2}) or from~(\ref{3.1}) follows eq.~(\ref{3.2}).

	Using the equation  $D_X=L_X+S_X$, one finds the following
 representations for the curvature and torsion operators:
	\begin{eqnarray*}
 R^D(X,Y) &=&
 S_X\circ S_Y - S_Y\circ S_X +
 [X,S_Y\cdot] - [Y,S_X\cdot]
  \> + \\
 & + & \>
 S_X([Y,\cdot]) - S_Y([X,\cdot]) - S_{[X,Y]} ,  \\
 T^D(X,Y) &=& S_X (Y) - S_Y (X) + [X,Y].
 	\end{eqnarray*}

	We have for them the following the local expressions:
	\begin{eqnarray}	\label{B.3}
\left[ \left( R^D(X,Y) \right)_{j}^{i} \right]  &=&
X(W_Y) - Y(W_X) + W_XW_Y - W_YW_X - W_{[X,Y]},	 \ \ \ \ \ \ \>    \\
\label{B.4}
\left( T^D(X,Y) \right)^{i}  &=&
\left(W_X\right)_{j}^{i} Y^j  -  \left(W_Y\right)_{j}^{i} X^j  -
C_{jk}^{i}X^jY^k,
	\end{eqnarray}
respectively

	For a linear connection $\nabla$ is fulfilled
 $\left(R^\nabla(X,Y)\right)_{j}^{i} = R_{jkl}^{i} X^k Y^l$  and
 $\left(T^\nabla(X,Y)\right)^i = T_{kl}^{i} X^k Y^l$ where
 $R_{jkl}^{i}$ and  $T_{kl}^{i}$  are the components of the usual curvature
and torsion tensors respectively~\cite{K&N-1,Schouten/Ricci}.

	Other general results concerning derivations can be found
in~\cite{K&N-1}.

 \section
[\hspace*{11.5ex} Sketch of some proofs]
{\bfseries Sketch of some proofs}
\label{C}
\setcounter{equation} {0}

	Propositions~\ref{Prop3.1}--\ref{Prop3.8} are the strict mathematical
basis for our analysis of the equivalence principle. Their full proofs can be
found
in~\cite{f-Frames-n+point,f-Bases-n+point,f-Bases-path,f-Bases-general}.
Below are presented the main aspects of them.

	\textit{Proof of proposition~\ref{Prop3.7}.}
	Let $\{ E^{\prime}_{i}=A_{i}^{j}E_j \}$ be a normal frame for $D$  in
$U$. Then $\left. W_{X}^{\prime}\right|_U=0$ which by~(\ref{3.2}) is
equivalent to
$W_X(x)=\Gamma_k(x)X^k(x),\ x\in U$  with
$\Gamma_k=-(E_k(A))A^{-1},\ A=[A_{i}^{j}]$.~\QED

	The first parts (necessity) of propositions~\ref{Prop3.1},
\ref{Prop3.3}, and~\ref{Prop3.5} are corollaries from
proposition~\ref{Prop3.7} when $U$  is a neighborhood, or a point, or a
path respectively. (Note that in the first case $W_X=-(X(A))A^{-1}$
implies $R^D=0$ due to~(\ref{B.3}).)

	\textit{Proof of proposition~\ref{Prop3.1} (sufficiency).}
	For a flat linear connection one can construct normal bases by
fixing some basis at an arbitrary point and then transporting it to any point
of  $U$  by means of the parallel transport generated by that
connection.~\QED

	\textit{Proof of proposition~\ref{Prop3.3} (sufficiency).}
	A local holonomic frame
$\{ E_{i}^{\prime}=A_{i}^{j}\partial/\partial x^j \}$
at a point $x_0$  can be constructed by choosing the coordinates $\{x^i\}$
such that $X=\partial/\partial x^1$ ($\neq 0$ at $x_0$) and putting
 $A(z) = \openone + C_k (x^k(z)-x^k(x_0))$
where $\openone$ is the unit matrix and
the matrices $C_k$  are partially fixed through the conditions
 $(C_k)_{j}^{i}=(C_j)_{k}^{i}\in\mathbb{R}$ and $C_1=W_X$.~\QED

	\textit{Proof of proposition~\ref{Prop3.5} (sufficiency).}
	Let the path $\gamma: J\to M$ be without self-intersections and be
contained in only one coordinate neighborhood.
Let $V:=J\times \cdots \times J$, where $J$ is taken $n-1$ times.  Let  us
fix a one-to-one $C^{1}$ map $\eta :J\times V\to M$ such  that
$\eta (\cdot ,\mathbf{t}_{0})=\gamma $  for  some fixed
$\mathbf{t}_{0}\in V$, i.e. $\eta (s,\mathbf{t}_{0})=\gamma (s), s\in J$.
(This is  possible  iff $\gamma $  is without self-intersections.) In
$U\bigcap \eta (J,V)$  we  introduce coordinates $\{x^{i}\}$ by putting
$(x^{1}(\eta (s,\mathbf{t})),\ldots ,x^{n}(\eta (s,\mathbf{t}))) =
(s,\mathbf{t})$, $s\in J, \mathbf{t}\in  V$.
(This, again, is possible iff $\gamma $ is without self-intersections.)
Let
 $W_X(\gamma(s))=\Gamma_k(\gamma(s)) X^k(\gamma(s)),\ s\in J$. Then all
normal along $\gamma$ frames
$\{ E_{i}^{\prime}=A_{i}^{j}\partial/\partial x^j \}$
are described by the matrix
	\begin{eqnarray}	\nonumber
A(\eta (s,\mathbf{t})) =
\left\{
\openone - \sum^{n}_{k=2}\Gamma _{k}(\gamma (s))
[x^{k}(\eta (s,\mathbf{t}))-x^{k}(\eta (s,\mathbf{t}_{0}))]
\right\} &\times&
\\  \nonumber \times \
Y(s,s_{0};-\Gamma _{1}\circ \gamma )B(s_{0},\mathbf{t}_{0};\eta ) &+&
\\ \label{C.1} + \
B_{kl}(s,\mathbf{t};\eta )
[x^{k}(\eta (s,\mathbf{t}))-x^{k}(\eta (s,\mathbf{t}_{0}))]
[x^{l}(\eta (s,\mathbf{t}))-x^{l}(\eta (s,\mathbf{t}_{0}))]. & &
	\end{eqnarray}
Here ${\openone}$ is the unit matrix, $s_{0}\in J$ is fixed, $B$ is  any
nondegenerate matrix function of its arguments, the matrix functions $B_{kl}$
are such that they and their first derivatives  are bounded when
$\mathbf{t}\to \mathbf{t}_{0}$, and $Y=Y(s,s_{0};Z)$, with $Z$ being a
continuous matrix function of $s$, is the unique solution  of
the matrix initial-value problem~\cite[ch.~IV, \S~1]{Hartman}
\[
{dY\over ds} =ZY, \quad \left.Y\right|_{s=s_{0}}={\openone}, \quad
Y=Y(s,s_{0};Z).
\]

	In the case when $\gamma$  has self-intersections and/or is not
contained in only one coordinate neighborhood the frames normal along
$\gamma$ are constructed from the ones for the pieces of $\gamma$  satisfying
the conditions at the beginning of this proof.~\QED

	\textit{Proof of proposition~\ref{Prop3.8} (first part).}
	If $\{E_i\}$  and  $\{E_i^\prime=A_i^j E_j\}$ are normal in $U$, then
 $\left. W_X \right|_U = \left. W_{X}^{\prime} \right|_U = 0$, which
by~(\ref{3.2}) means that $\left.X(A)\right|_U=0$, i.e.
$\left. E_i(A)\right|_U = 0$ as  $X$  is arbitrary. Conversely, if $\{E_i\}$
is normal in $U$, i.e. $\left. W_X \right|_U = 0$, and $E_i^\prime=A_i^j E_j$
with $\left. E_i(A)\right|_U = 0$, then, again by~(\ref{3.2}), we get
 $\left. W_{X}^{\prime} \right|_U = 0$, i.e. $\{E_{i}^{\prime}\}$  is normal
in~$U$.~\QED

	If we specify $U$  to be neighborhood, or a point, or a curve (i.e.
the set $\gamma(J))$, then from the first part of proposition~\ref{Prop3.8}
follow the first parts of propositions~\ref{Prop3.2}, \ref{Prop3.4},
and~\ref{Prop3.6} respectively. Analogously, their second parts are
corollaries from the second part of proposition~\ref{Prop3.8}.

	\textit{Proof of proposition~\ref{Prop3.8} (second part).}
	If $\{E_{i}^{\prime}\}$ is a normal frame in $U$, then
 $\left. W_{X}^{\prime} \right|_U = 0$ which, due to~(\ref{B.4}), implies
\(
 \left. T^D ( E_{i}^{\prime},E_{j}^{\prime} ) \right|_U =
- \left. [ E_{i}^{\prime},E_{j}^{\prime} ] \right|_U.
\)
So, the holonomicity condition
$\left.[(E_{i}^{\prime},E_{j}^{\prime}]\right|_U=0$
is equivalent to $\left.T^D\right|_U=0$.~\QED

	The considered propositions can be proved also independently,
which is done in the above-cited references, where other details and
results can be found.

\bibliography{bozhopub,bozhoref}
\bibliographystyle{unsrt}

\end{document}